\documentclass[a4paper,11pt]{article}
\usepackage{slashed}
\usepackage{pos}

\title{The strong CP puzzle and axions}
\author*[a]{Christopher Smith}

\affiliation[a]{Laboratoire de Physique Subatomique et de Cosmologie,
 \\ Universit\'{e} Grenoble-Alpes, CNRS/IN2P3, Grenoble INP, 38000 Grenoble, France}

\emailAdd{chsmith@lpsc.in2p3.fr}

\abstract{In the first part of this talk, after a brief presentation of the strong CP puzzle, the construction of axion models and their main phenomenological features are described. In the second part, the possibility to mix the Peccei-Quinn symmetry with baryon and lepton numbers is discussed, showing that the axion could ultimately play a role in other puzzles of the Standard Model like the smallness of neutrino masses or baryogenesis.}

\FullConference{31st International Workshop on Deep Inelastic Scattering (DIS2024)\\
 8–12 April 2024\\
Grenoble, France\\}

\begin{document}
\maketitle

\section{Brief introduction to the strong CP puzzle}

The Standard Model (SM) is defined as containing all the possible renormalizable couplings compatible with its gauge symmetries and matter content. This should include the CP-violating%
\begin{equation}
\mathcal{L}_{SM}\supset\frac{\alpha_{S}}{4\pi}\theta\operatorname*{tr}G_{\mu\nu}\tilde{G}^{\mu\nu}\ , 
\label{Eq1}
\end{equation}
where $\tilde{G}^{\mu\nu}=\varepsilon^{\mu\nu\rho\sigma}G_{\rho\sigma}/2$ is the dual of the gluon field strength tensor, and $\theta$ an additional free parameter. Such a coupling induces a CP-violating term to the neutron coupling to photons, $\langle\gamma n|\operatorname*{tr}G_{\mu\nu}\tilde{G}^{\mu\nu}|n\rangle$, encoded at leading order in the non-relativistic expansion into its electric dipole moment (EDM). Experimentally, no such EDM has been found~\cite{Abel:2020pzs}, and the current limit $\theta<10^{-10}$ appears at odd with naturality. Yet, the problem is actually far more serious because, in the SM, two unrelated mechanisms contribute to $\theta$.

\begin{figure}[t]
\centering\includegraphics[width=0.70\textwidth]{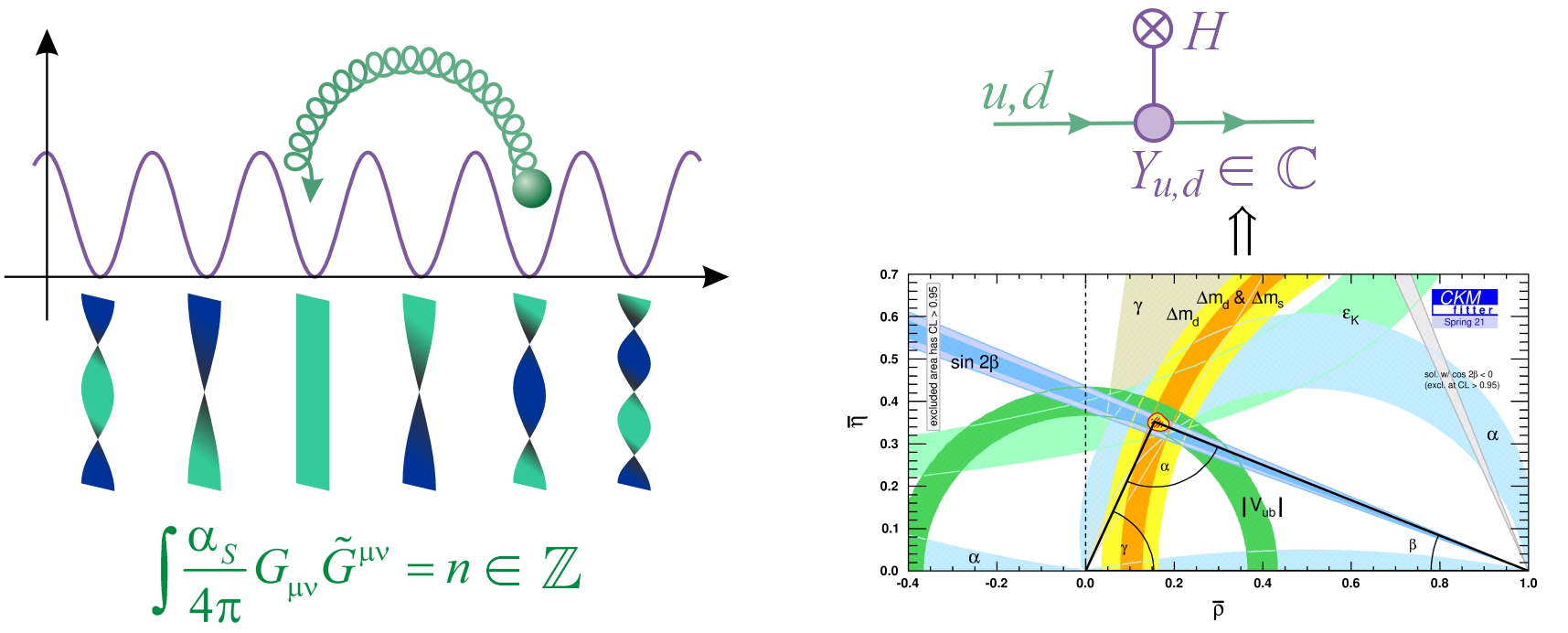}
\caption{Illustration of the two sources for $\theta$ in the SM. In this
picture, a naive way to see that the QCD topology breaks $T$, hence $CP$, is
to imagine a gluon emitted from the vacuum at time $t_{1}$ and absorbed at a
later time $t_{2}$. The time-reversed process is different since emitting back
the gluon at $t_{2}$, there is no guarantee to end up in the same vacuum at
$t_{1}$.}
\label{Fig1}
\end{figure}

First, there is a strong interaction contribution of topological origin, $\theta_{glue}$. That it must be topological is understood from the peculiar nature of the coupling of Eq.~(\ref{Eq1}): it is a total derivative. Thus, only boundary terms could make it non-zero. Usually, fields are required to vanish asymptotically, but in the presence of a gauge symmetry, one should rather ask them to be pure gauge at infinity. This is where the topology of the color group $SU(3)$ comes into play: vacuum gauge configurations fall into equivalence classes according to how many times they wind over the sphere at infinity, see Fig.~\ref{Fig1}. This would be rather innocuous but for the existence of field configurations, the instantons, tunneling between these gauge configurations. Thus, the true vacuum of QCD is a specific superposition, labelled by $\theta_{glue}$, of these topologically distinct vacua. Importantly, these mathematical considerations are not pure speculations but do have experimental support in the mass of the $\eta^{\prime}$. This state can mix with $\operatorname*{tr}G_{\mu\nu}\tilde{G}^{\mu\nu}$, now interpreted as a glueball made heavy by the presence of instantons~\cite{eta}. Numerically, this is measured on the lattice~\cite{Bali:2021qem} via the topological susceptibility $\chi\sim\langle0|(\operatorname*{tr}G_{\mu\nu}\tilde{G}^{\mu\nu})(\operatorname*{tr}G_{\mu\nu}\tilde{G}^{\mu\nu})|0\rangle$. Thus, a priori, $\theta_{glue}$ should be an $\mathcal{O}(1)$ parameter.

A second contribution to $\theta$ comes from the Higgs Yukawa coupling to quarks, $\mathbf{Y}_{u,d}$. Those are complex matrices, hence after symmetry breaking, anomalous rephasing of the quark fields are required to make the quark masses real, generating $\theta_{flav}=\arg\det\mathbf{Y}_{u}+\arg\det\mathbf{Y}_{d}$. Here also, we have experimental support for $\mathbf{Y}_{u,d}$ being complex since weak CP violation from the CKM matrix is well-established~\cite{Charles:2004jd}, see Fig.~\ref{Fig1}. Thus, $\theta_{flav}$ is also a free $\mathcal{O}(1)$ parameter.%

Now, the non-observation of strong CP violation requires a near perfect cancellation, to one part in ten billion, between the QCD topological contribution and the Higgs Yukawa contributions, $\theta=\theta_{glue}+\theta_{flav}<10^{-10}$. Arguing that somehow QCD has to be defined on that specific vacuum where CP is conserved is clearly not sufficient since this only sets $\theta_{glue}$ to zero. As the Higgs and QCD sectors are entirely separated in the SM, with the Higgs boson not even colored, the situation is way more puzzling than having simply the Lagrangian parameter $\theta$ accidentally small. This mysterious cancellation is so puzzling that significant departures from the SM are envisioned, with the axion solution being the prime candidate. Yet, before turning to that, let us briefly discuss some alternative solutions that have appeared along the years.

\subsubsection*{Solution 1: Massless quark}

Historically, the first solution to the strong CP puzzle is also the simplest.\ If one quark flavor is massless, say the up quark, then its right and left components can be rephased independently. But any mismatch between the rephasing of these two chiralities generates a shift in $\theta$ thanks to the chiral anomaly. Thus, $\theta$ ceases to be physical and can be rotated to zero.

Nowadays, chiral perturbation theory combined with lattice QCD forbid this solution, with the smallest quark mass $m_{u}$ definitely greater than zero~\cite{Leutwyler:2009jg}. Yet, it should be remarked that what matters to solve the strong CP puzzle is the absence of a Lagrangian mass term for the up quark, whose physical mass also receives instanton contributions~\cite{Georgi:1981be}. Though QCD alone cannot suffice, if one can concoct a model in which some instanton contributions do saturate the measured up-quark mass, then QCD becomes CP conserving again (see e.g. Ref.~\cite{Carena:2019nnd} and references there).

\subsubsection*{Solution 2: GUT paradigm}

Per se, Grand Unified Theories (GUT) do not solve the strong CP puzzle, but they may alleviate it somehow. First, some GUT theories can force Yukawa matrices to be hermitian. This is typically the case when a left-right symmetry is active~\cite{Senjanovic:2020int}. At the GUT scale, this ensures $\theta_{flav}=0$. Of course, since CP is violated in the electroweak sector, this cannot be the end of the story~\cite{RunTheta}. Finite contributions from the CKM phase are tiny though, below $10^{-16}$. Infinite contributions, arising first at seven loops, make $\theta$ a running parameter. Those sum up to even tinier corrections below $10^{-18}$. All that remains thus is the issue of $\theta_{glue}$. Inspired by how the electroweak $\theta$ term gets rotated away in the SM thanks to the residual but anomalous $U(1)$ symmetry corresponding to baryon plus lepton numbers, $\mathcal{B}+\mathcal{L}$, a similar mechanism may be invoked~\cite{Shifman:2017lkj}. To our knowledge, a full model with all these ingredients, able to account for the unification of the couplings and realistic fermion masses has not been proposed yet.

\subsubsection*{Solution 3: Infinities}

The reality of the strong CP puzzle rests on how to deal with no less than three infinities. First, the impact of $\operatorname*{tr}G_{\mu\nu}\tilde{G}^{\mu\nu}$ vanishes at any finite order of perturbation theory. Second, being a total derivative, it is the boundary at infinity that matters. Third, the vacuum $|\theta\rangle$ is an infinite sum over the topologically distinct sectors $|k\rangle$, i.e., $|\theta\rangle=\sum_{k}e^{ik\theta}|k\rangle$.

Messing up with how these infinities are treated individually or in conjunction with one another can drastically change the outcome. The main constraint here is to make sure $\operatorname*{tr}G_{\mu\nu}\tilde{G}^{\mu\nu}$ does contribute to the $\eta^{\prime}$ mass. Another difficulty to design a paradigm within QCD to drive $\theta$ to zero is that QCD with a finite $\theta$ probably does not look much like the QCD we know, since it may not even be confining~\cite{tHooft:1981bkw}. Thus, phenomenology does not appear to behave smoothly as a function of $\theta$. Finally, lattice simulations are not always of great help. While the topological susceptibility, and thereby the $\eta^{\prime}$ mass, are well reproduced, the impact of the $\theta$ term is more elusive, with e.g. the neutron EDM matrix element being still compatible with zero~\cite{Lattice}.

Recently, the focus has been on the order of the infinite limits~\cite{Ai:2020ptm} for the volume over which correlation functions are defined and the summation over $k$. Sketchily, the idea is that if the volume is sent to infinity before the sum over $k$, then each $k$ region becomes infinitely large, and all our observables are forced to live in a given vacuum, see Fig.~\ref{Fig1}. In our opinion, whether such a non-standard reversal in the order of the limits is permitted is not established though. So, for the time being, our point of view will be that the strong CP puzzle appears quite resilient, but given the complexity of the QCD dynamics, one cannot rule out the possibility that it resolves itself spontaneously in the future.

\section{Axions to solve the strong CP puzzle}

The main ingredient of the axion solution to the strong CP puzzle, along with its phenomenology, can be illustrated starting from:%
\begin{equation}
\mathcal{L}\supset\frac{\alpha_{S}}{4\pi}\theta\operatorname*{tr}G_{\mu\nu
}\tilde{G}^{\mu\nu}+(y \phi\bar{\psi}_{L}\psi_{R}+h.c.)+\partial_{\mu}\phi
^{\dagger}\partial^{\mu}\phi-V(\phi^{\dagger}\phi)\ . 
\label{Eq21}
\end{equation}
The complex scalar field $\phi$ does not belong to the SM, though the colored fermion $\psi$ may.

\subsubsection*{The Peccei-Quinn step:}

The first step is to enforce the $U(1)_{PQ}$ symmetry generated by $\phi\rightarrow\exp(i\theta)\phi$ by giving different charges to $\psi_{L}$ and $\psi_{R}$. The associated Noether current then has an anomaly, $\partial_{\mu}J_{PQ}^{\mu}\sim\operatorname*{tr}G_{\mu\nu}\tilde{G}^{\mu\nu}$. At the same time, $V(\phi^{\dagger}\phi)$ is assumed to break $U(1)_{PQ}$ spontaneously, so the scalar field can be represented as
\begin{equation}
\phi(x)=(f_{a}+\rho(x))\exp(ia(x)/f_{a})\ , 
\label{Eq22}
\end{equation}
where $f_{a}$ is the vacuum expectation value and the axion $a$ is the Goldstone boson whose shifts span the circular vacuum. By Goldstone theorem, $a$ is coupled to $\operatorname*{tr}G_{\mu\nu}\tilde{G}^{\mu\nu}$ since the current satisfies $\langle0|J_{PQ}^{\mu}|a(p)\rangle=if_{a}p^{\mu}$. The $\rho$ field has a mass of $\mathcal{O}(f_{a})$, so integrating it out, the Lagrangian at low energy is generically%
\begin{equation}
\mathcal{L}\supset\frac{1}{2}\partial_{\mu}a\partial^{\mu}a+\frac{\alpha_{S}%
}{4\pi}\left(  \theta+\frac{a}{f_{a}}\right)  \operatorname*{tr}G_{\mu\nu
}\tilde{G}^{\mu\nu}+\frac{1}{f_{a}}\partial_{\mu}aJ_{PQ}^{\mu}+...\ .
\label{Eq23}
\end{equation}

\subsubsection*{The invisibility step:}

\begin{figure}[t]
\centering\includegraphics[width=0.90\textwidth]{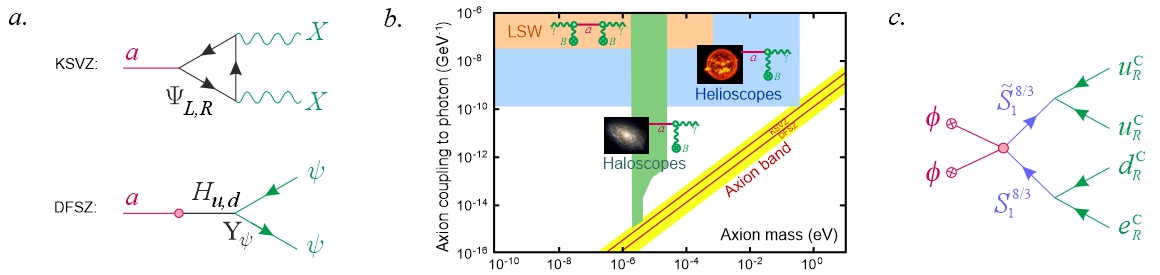}
\caption{$a.$ Leading axion couplings to SM particles in the KSVZ and DFSZ
scenarios. $b.$ Schematic representation of the axion exclusion plot, with the
axion band from the theoretical correlation between $g_{a\gamma\gamma}$ and
$m_{a}$. The main search strategies are light-shining through the wall-type
experiments (LSW), helioscopes using the sun as an axion source, or haloscopes
aiming for the DM axions of the galactic halo. $c.$ Spontaneous proton decay
in the model of Eq.~(\ref{LQ}).}
\label{Fig2}
\end{figure}

Whether it is via $\operatorname*{tr}G_{\mu\nu}\tilde{G}^{\mu\nu}$, itself coupled to $\eta^{\prime}$, or via SM fermions present in $J_{PQ}^{\mu}$, the axion is coupled to normal matter. Its non-observation immediately requires $f_{a}$ to be well above the electroweak scale $v_{EW}$. There are then essentially two paradigms to make the axion "invisible". The KSVZ scenario~\cite{KSVZ} introduces both $\phi$ and $\psi_{L,R}$ as new degrees of freedom. To preserve the renormalizability of the SM, $\psi_{L,R}$ are then vectorlike under all the SM symmetries. Generically, the Lagrangian at the electroweak scale is then (see Fig.~\ref{Fig2})
\begin{equation}
\mathcal{L}\supset\frac{a}{16\pi^{2}f_{a}}\sum_{X=G,W,B}g_{X}^{2}N_{X}X_{\mu\nu}\tilde{X}^{\mu\nu}\ , 
\label{Eq24}
\end{equation}
with $N_{X}$ functions of the SM gauge charges of $\psi$.

The other scenario starts from a two-Higgs doublet model (THDM)~\cite{PQ}. For a type II fermionic sector, and provided the scalar potential leaves room for the $U(1)_{PQ}$ symmetry, the electroweak symmetry breaking makes the pseudoscalar Higgs boson massless~\cite{Axion}, so $a=A^{0}$ but $f_{a}\sim v_{EW}$. To make an invisible axion, the DFSZ idea~\cite{DFSZ} is to add the complex scalar field of Eq.~(\ref{Eq22}) together with a coupling $\phi^{2}H_{u}H_{d}$, such that the physical axion field becomes $a\rightarrow a+(v_{ew}/f_{a})A^{0}$. All the couplings of $a$ to SM fields come from that of $A^{0}$, see Fig.~\ref{Fig2}, hence to leading order,%
\begin{equation}
\mathcal{L}\supset\frac{a}{f_{a}}\sum_{\psi=u,d,e}\chi_{\psi}m_{\psi}\bar{\psi}\gamma^{5}\psi\ , 
\label{Eq25}
\end{equation}
where $\chi_{\psi}$ depend on the vacuum expectation values of $H_{u,d}$. Of course, at the loop level, both the KSVZ and DFSZ include couplings to fermions and gauge bosons.

\subsubsection*{The hadronic step:}

No matter the precise implementation, the outcome is a massless field coupled
to $\operatorname*{tr}G_{\mu\nu}\tilde{G}^{\mu\nu}$, and possibly also to
$F_{\mu\nu}\tilde{F}^{\mu\nu}$:%
\begin{equation}
\mathcal{L}\supset\frac{\alpha_{S}}{4\pi}\left(  \theta+\frac{a}{f_{a}%
}\right)  \operatorname*{tr}G_{\mu\nu}\tilde{G}^{\mu\nu}+\frac{1}{4}%
g_{a\gamma\gamma}^{SD}aF_{\mu\nu}\tilde{F}^{\mu\nu}+...\ . \label{Eq26}%
\end{equation}
The $\theta$ term then necessarily relaxes to zero. Indeed, once QCD hadronization sets in, the circular vacuum space of the potential gets tilted, and $a$ falls to a specific vacuum which happens to be that where CP is conserved. Technically, $\phi\operatorname*{tr}G_{\mu\nu}\tilde{G}^{\mu\nu}$ becomes a potential $V(\phi)$, whose minimum is at $\langle\phi\rangle=0$. A recurrent difficulty here is that this tilting is not that strong, so the PQ symmetry needs to be near perfect. This is the axion quality problem: even Planck-scale tilting of the potential could jeopardize the final drive to the 
CP-conserving solution~\cite{Kamionkowski:1992mf}.

There are other important consequences (see e.g. Ref.~\cite{DiLuzio:2020wdo} for a review). First, $a$ becomes massive since the vacuum space is
no longer flat, with $f_{a}m_{a}\approx f_{\pi}m_{\pi}$. Second, $a$ mixes with the light neutral mesons $\pi^{0}$, $\eta$, and $\eta^{\prime}$ since $\operatorname*{tr}G_{\mu\nu}\tilde{G}^{\mu\nu}$ mixes with the singlet $\eta_{0}$. This immediately pushes $f_{a}$ to very large values, e.g. from $K\rightarrow\pi(\pi\rightarrow a)$. Third, even if in a specific model, $g_{a\gamma\gamma}^{SD}=0$, the axion inherits a $\gamma\gamma$ coupling $g_{a\gamma\gamma}^{LD}$ from that of $\pi^{0}$, $\eta$, and $\eta^{\prime}$. Barring fine-tuning, $g_{a\gamma\gamma}=g_{a\gamma\gamma}^{SD}+g_{a\gamma\gamma}^{LD}$ is then correlated to the mass, $g_{a\gamma\gamma}/m_{a}\approx10^{-1\pm1}$, see Fig.~\ref{Fig2}. Fourth, the axion is sufficiently long-lived to become a cold dark matter (DM) candidate. If the axion field is initially misaligned with the CP-conserving vacuum, its falling to it corresponds to collective oscillations whose zero-mode energy density can match the observed dark matter abundance. With $m_{a}$ related to the topological susceptibility, and from lattice estimates of the temperature dependence of the latter, $10-1000~\mu eV$ emerges as a preferred $m_a$ range for $\mathcal{O}(1)$ initial misalignment angles.%

\subsubsection*{The discovery step?}

All the ingredients are now in place. The axion has its characteristic
coupling to $F_{\mu\nu}\tilde{F}^{\mu\nu}=E\cdot B$, correlated to its mass.
Search strategies exploit this coupling mostly as an axion-photon conversion in
an intense $B$ field, and their constraints are then represented in the
$g_{a\gamma\gamma}$ vs. $m_{a}$ plane, see Fig.~\ref{Fig2}. There are many
such experiments currently going on or planned in the near future (including a
local initiative in Grenoble~\cite{Grenet:2021vbb}), along with many efforts
aiming at the axion couplings to leptons or nucleons (see \cite{AxionExp} for
an up-to-date list and corresponding exclusion plots). It should be said also
that Fig.~\ref{Fig2} represents the 'vanilla' scenarios, but many alternatives
exist, with models out of the axion band (using larger
representations, clockwork, some $Z_{n}$ symmetry, mirror QCD,..., see e.g. Refs.~\cite{Outside}), or out of the preferred DM mass range (with some kinetic
energy~\cite{Co:2019jts} or topological defects~\cite{Marsh:2015xka}).

\section{Baryonic axions}

Introducing the axion to solve the strong CP puzzle is a rather high price to
pay, not least because one is forced to add also new fermions (like in KSVZ),
scalars (like in DFSZ), or even a whole new dark sector. Its capability to
explain the observed dark matter relic density strengthen its case, but in a
spirit of minimality, it would be desirable to further its phenomenological
role. This is where the interplay with baryon and lepton numbers,
$\mathcal{B}$ and $\mathcal{L}$, could prove fruitful. Indeed, at the root of
the axion mechanism is the PQ symmetry, under which some colored fermions must
transform non-trivially. As such, this symmetry is nothing more than a flavor
$U(1)$ symmetry, exactly like $\mathcal{B}$ and $\mathcal{L}$. In practice,
these symmetries naturally mix, as we will see. This opens the way for the
axion to play a role in the generation of neutrino masses and leptogenesis, or
even directly in baryogenesis. Given the accidental closeness of the DM and
baryonic relic densities, having a model relating both would certainly look
promising. This is still a long way off. In the present section, the goal is
to summarize some recent advances made incorporating $\mathcal{B}$ and
$\mathcal{L}$ violation within axion models.

\subsubsection*{Reparametrization freedom}

Before embarking into model building, it is crucial to realize that axion
models can be defined in very different though equivalent forms. This is
well-known in the context of Chiral Perturbation Theory, but not always fully
appreciated for axions. Specifically, starting from the renormalizable
Lagrangian of Eq.~(\ref{Eq21}), the usual linear representation is%
\begin{equation}
\phi\sim f_{a}+\rho+ia:\mathcal{L}_{linear}=\frac{\alpha_{S}\theta}{4\pi
}\operatorname*{tr}G_{\mu\nu}\tilde{G}^{\mu\nu}+\bar{\psi}(i\slashed{D}-m)\psi
-im\frac{a}{f_{a}}\bar{\psi}\gamma^{5}\psi+...\ .
\end{equation}
The Lagrangian stays renormalizable after symmetry breaking, with the axion
having pseudoscalar couplings to fermions. Importantly, no process is
anomalous in this description, even at loop level. This is evident in the
original Peccei-Quinn model since it is simply matched onto the THDM.

In view of the topology of the vacuum (i.e., a circle), one can instead adopt
the polar representation for $\phi$. The Lagrangian is then no longer
manifestly renormalizable%
\begin{equation}
\phi\sim(f_{a}+\rho)\exp(ia/f_{a}):\mathcal{L}_{polar}=\frac{\alpha_{S}\theta
}{4\pi}\operatorname*{tr}G_{\mu\nu}\tilde{G}^{\mu\nu}+\bar{\psi}%
(i\slashed{D}-m\exp(ia\gamma^{5}/f_{a}))\psi+...\ .
\end{equation}
The interest of that representation is that $\rho$ decouples and can trivially
be integrated out. Also, it shows explicitly how the Goldstone boson makes the
fermion mass term compatible with the chiral $U(1)_{PQ}$ symmetry.

In the previous representation, a $U(1)_{PQ}$ transformation changes
$\theta\rightarrow\theta+\alpha$ by first shifting $a\rightarrow a+f_{a}%
\alpha$, which then requires a compensating chiral, hence anomalous, rotation
of the fermion field $\psi\rightarrow\exp(i\alpha/2)\psi$. It is possible to
skip this last step by adopting a reparametrization of the fermion field to
make it neutral under $U(1)_{PQ}$ (compare with Eq.~(\ref{Eq23})):%
\begin{equation}
\psi\rightarrow\exp(-ia\gamma^{5}/(2f_{a}))\psi:\mathcal{L}_{derivative}%
=\frac{\alpha_{S}}{4\pi}\left(  \theta+\frac{a}{v}\right)  \operatorname*{tr}%
G_{\mu\nu}\tilde{G}^{\mu\nu}+\bar{\psi}\left(  i\slashed{D}-m+\frac{\partial_{\mu
}a\gamma^{\mu}\gamma^{5}}{f_{a}}\right)  \psi+...\ .
\end{equation}
This is the derivative representation, in which the axion appears essentially
as a dynamical $\theta$ term, but with derivative interactions otherwise, as
appropriate for a Goldstone boson.

Though Feynman rules, diagrams, symmetry properties, and non-relativistic
expansions~\cite{Smith:2023htu} are different in the three representations
$\mathcal{L}_{linear}$, $\mathcal{L}_{polar}$, and $\mathcal{L}_{derivative}$,
observables have to be the same whatever the chosen form. In most
applications, the derivative representation is taken as a starting point. For
most low-energy applications, this is perfectly fine, but there are issues
when $\mathcal{B}$ and/or $\mathcal{L}$ violation is present, or when weak
interactions are considered. Let us discuss these points in turns.

\subsubsection*{Electroweak couplings}

To illustrate the issue with the weak interaction~\cite{Blind}, let us adopt
in a simplified setting with only a single fermion, as in Eq.~(\ref{Eq21}).
Starting from a Yukawa term like $y\phi\bar{\psi}_{L}\psi_{R}$, the PQ charges
are only defined up to a common constant. Setting $Q_{\phi}=1$, we must have
$Q_{\psi_{L}}=\alpha$ and $Q_{\psi_{R}}=\alpha-1$. The free parameter $\alpha$
reflects the existence of another $U(1)$ flavor symmetry, corresponding to the
conservation of the $\psi$-number. Now, if $\psi$ is charged under some gauge interaction $X$, the effective axion Lagrangian can be constructed as%
\begin{align}
\mathcal{L}_{derivative}  &  \supset\frac{a}{f_{a}}\frac{g_{X}^{2}}{16\pi^{2}%
}(Q_{\psi_{L}}C_{2}^{\psi_{L}}-Q_{\psi_{R}}C_{2}^{\psi_{R}})\operatorname*{tr}%
X_{\mu\nu}\tilde{X}^{\mu\nu}\nonumber\\
&  +\frac{\partial_{\mu}a}{f_{a}}\left(  (Q_{\psi_{R}}+Q_{\psi_{L}})\bar{\psi
}\gamma^{\mu}\psi+(Q_{\psi_{R}}-Q_{\psi_{L}})\bar{\psi}\gamma^{\mu}\gamma
^{5}\psi\right)  \ .
\end{align}
For QCD and QED, the Casimir invariant $C_{2}^{\psi_{L}}=C_{2}^{\psi_{R}}$,
hence $\alpha$ drops out of the anomalous term. It also disappears from the
axial current, but remains in the vector one. However, that current being
conserved, it cannot contribute to observables. Hence, the $\psi$-number
ambiguity is irrelevant.

The situation changes for a chiral interaction, for which $C_{2}^{\psi_{L}%
}\neq C_{2}^{\psi_{R}}$. For example, if $\psi_{L}$ is a weak doublet, but
$\psi_{R}$ a singlet, then $aW_{\mu\nu}\tilde{W}^{\mu\nu}$ is entirely tuned
by the free parameter $\alpha$. This reflects the fact that the $W_{\mu\nu
}\tilde{W}^{\mu\nu}$ term comes entirely from the anomaly in the $\psi$-number
symmetry current. The $aW_{\mu\nu}\tilde{W}^{\mu\nu}$ coupling then depends on
how much of that current is present in the PQ current. One could think of
setting $\alpha=0$ to somewhat remove the $\psi$-number current, but this
totally kills the $aW_{\mu\nu}\tilde{W}^{\mu\nu}$ coupling, which is not
consistent either since clearly, $a\rightarrow W^{+}W^{-}$ or $a\rightarrow
ZZ$ can occur if $\psi_{L}$ has weak interactions, see Fig.~\ref{Fig2}.

Actually, this dependence on $\alpha$ is spurious, because the local anomalous terms $aW_{\mu\nu}\tilde{W}^{\mu\nu}$ and $aB_{\mu\nu}\tilde{B}^{\mu\nu}$ in $\mathcal{L}_{derivative}$ are spurious. The reason is that in the presence of chiral gauge interactions, neither the axial nor the vector current are conserved: the triangle graphs involving both these currents are anomalous. Crucially, the anomalous terms from these triangle graphs exactly cancel with the local anomalous terms. This is expected since we could instead work in the linear representation, in which no anomaly ever arises. In other words, anomalies are an artifact of the derivative representations. At the end of the day, the axion couplings thus come from the non-anomalous part of the triangle graphs, and precisely match the result one would get starting from the pseudoscalar interaction in either $\mathcal{L}_{linear}$ or $\mathcal{L}_{polar}$. What makes the situation a bit peculiar for QED and QCD is that this non-anomalous remainder is \textit{parametrically} identical, in the $m\rightarrow\infty$ limit, to the anomalous term (this is the Sutherland-Veltman theorem). But this does not work for chiral gauge bosons.

\subsubsection*{Merging the PQ symmetry with $\mathcal{B}$ and $\mathcal{L}$}

In realistic models, beside the $\psi$-number discussed above for KSVZ-type
models, there are two additional ambiguities due to $\mathcal{B}$ and
$\mathcal{L}$, since those are always conserved by the SM Yukawa couplings. It
is important to keep track of these ambiguities whenever $\mathcal{B}$ and/or
$\mathcal{L}$ violating interactions are turned on~\cite{BandL}.

Let us give a simple example. We add to the KSVZ model the coupling $\phi
\bar{\nu}_{R}^{\mathrm{C}}\nu_{R}+h.c.$ . This merges lepton number with the
PQ charges, because one is forced to set not only the PQ charge of $\nu_{R}$
to $-1/2$, but also that of the lepton weak doublet $\ell_{L}=(\nu_{eL}%
,e_{L})$ and singlet $e_{R}$ because of $\bar{e}_{R}Y_{e}\ell_{L}$ and
$\bar{\nu}_{R}Y_{\nu}\ell_{L}$. Notice that if one constructs $\mathcal{L}%
_{derivative}$ using these charges, spurious couplings of the axion to charged
leptons appear, showing once again how plagued by cancellations is that
representation. More importantly, had we frozen those PQ charges to zero, not
accounting for the ambiguity due to $\mathcal{L}$, one may have wrongly
concluded that the $\phi\bar{\nu}_{R}^{\mathrm{C}}\nu_{R}$ coupling is
incompatible with the PQ symmetry. This is obviously wrong since nothing
forbids it in the linear representation. What is true though is that whenever
$\phi$ is coupled to normal matter, it is impossible to incorporate a PQ
symmetry if more than two different combinations of $\mathcal{B}$ and
$\mathcal{L}$ are broken. For example, if $\phi\bar{\nu}_{R}^{\mathrm{C}}%
\nu_{R}$ is included to merge the seesaw and axion mechanisms, then only a
single other type of $\mathcal{B}$ and/or $\mathcal{L}$ violation is possible
before the axion becomes massive. In this context, one should also pay
attention to electroweak instantons, generating the effective $(\ell_{L}%
q_{L}^{3})^{3}$ interaction, that can induce axion quality
problems~\cite{BandL}.

To force $\phi$ to carry other combinations of $\mathcal{B}$ and $\mathcal{L}$
is not so easy though. Clearly, a direct coupling to three quarks (plus a
lepton) would not be renormalizable. The strategy systematically analyzed in
Ref.~\cite{Arias-Aragon:2022byr} is to do this via various types of
leptoquarks and diquarks. Then, one can merge the PQ symmetry with any two
combinations of $\mathcal{B}$ and $\mathcal{L}$ allowed by the gauge
symmetries. For example, if one adds two weak singlet scalar states of
hypercharge 8/3 and couplings%
\begin{equation}
\mathcal{L}=\mathcal{L}_{KSVZ}+\phi\bar{\nu}_{R}^{\mathrm{C}}\nu_{R}%
+S_{1}^{8/3}\bar{d}_{R}e_{R}^{\mathrm{C}}+\tilde{S}_{1}^{8/3}\bar{u}%
_{R}^{\mathrm{C}}u_{R}+\phi^{2}S_{1}^{8/3\dagger}\tilde{S}_{1}^{8/3}%
+h.c.\ ,\label{LQ}%
\end{equation}
then a single global $U(1)$ merging PQ, $\mathcal{B}$ and $\mathcal{L}$ remains.
This symmetry prevents any other couplings, and when $\phi$ breaks it
spontaneously, neutrinos become massive and proton decay arises, see
Fig.~\ref{Fig2}. Even though these models are hardly economical, they
illustrate how varied the phenomenological possibilities are, with features
like spontaneous proton decay, neutron-antineutron oscillations, neutrino
masses and neutrinoless double-beta decay, a possibility to account for the
neutron lifetime puzzle (for an ALP with a mass conveniently set right
in-between that of the proton and neutron), or with axion-induced
neutron-antineutron transitions.

\section{Conclusion}

Currently, the axion remains our best mechanism to solve the strong CP puzzle. Further, as the weakly-interacting massive particle paradigm suffers from the absence of signals of new physics at colliders, it is becoming one of our best candidates as constituent of the elusive dark matter. Though from a
model-building perspective, the axion mechanism is a rather simple departure from the SM, what is remarkable is that it could
also hold the key to other long-standing puzzles. For instance, the similarity in the PQ breaking and seesaw scale points towards a role of the axion in the smallness of neutrino masses, and that between baryonic and dark matter relic densities advocates for an axionic baryogenesis of some kind. With experimental efforts now closing in on the parameter space expected in the simplest scenarios, a discovery could be around the corner. This could revolutionize both particle physics and cosmology, and open the door to many further advances in our understanding of Nature.


\begin{thebibliography}{99}                                                                                               %

\bibitem {Abel:2020pzs}C.~Abel \textit{et al.}
%``Measurement of the Permanent Electric Dipole Moment of the Neutron,''
Phys. Rev. Lett. \textbf{124} (2020) no.8, 081803 [arxiv:2001.11966].
%336 citations counted in INSPIRE as of 30 Jan 2023

\bibitem {eta}G.~'t Hooft,
%``Symmetry Breaking Through Bell-Jackiw Anomalies,''
Phys. Rev. Lett. \textbf{37} (1976) 8;
%4197 citations counted in INSPIRE as of 12 Jul 2024
%``Computation of the Quantum Effects Due to a Four-Dimensional Pseudoparticle,''
Phys. Rev. D \textbf{14} (1976) 3432;
%4509 citations counted in INSPIRE as of 12 Jul 2024
E.~Witten,
%``Current Algebra Theorems for the U(1) Goldstone Boson,''
Nucl. Phys. B \textbf{156} (1979) 269.
%1703 citations counted in INSPIRE as of 12 Jul 2024

\bibitem {Bali:2021qem}G.~S.~Bali \textit{et al.} [RQCD],
%``Masses and decay constants of the \ensuremath{\eta} and \ensuremath{\eta}' mesons from lattice QCD,''
JHEP \textbf{08} (2021) 137 [2106.05398].
%46 citations counted in INSPIRE as of 12 Jul 2024

\bibitem {Charles:2004jd}J.~Charles \textit{et al.} [CKMfitter Group],
%``CP violation and the CKM matrix: Assessing the impact of the asymmetric $B$ factories,''
Eur. Phys. J. C \textbf{41} (2005) 1 [hep-ph/0406184], updated
results and plots available at: http://ckmfitter.in2p3.fr.
%2124 citations counted in INSPIRE as of 12 Jul 2024


\bibitem {Leutwyler:2009jg}H.~Leutwyler,
%``Light quark masses,''
PoS \textbf{CD09} (2009) 005 [0911.1416].
%42 citations counted in INSPIRE as of 05 Jul 2024


\bibitem {Georgi:1981be}H.~Georgi and I.~N.~McArthur,
%``INSTANTONS AND THE mu QUARK MASS,''
HUTP-81/A011.
%8 citations counted in INSPIRE as of 05 Jul 2024


\bibitem {Carena:2019nnd}M.~Carena, \textit{et al.}
%``$\nu$ solution to the strong CP problem,''
Phys. Rev. D \textbf{100} (2019) 9, 094018 [1904.05360].
%12 citations counted in INSPIRE as of 05 Jul 2024


\bibitem {Senjanovic:2020int}G.~Senjanovic and V.~Tello,
%``Strong CP violation: Problem or blessing?,''
Int. J. Mod. Phys. A \textbf{38} (2023) 15n16, 2350067 [2004.04036].
%17 citations counted in INSPIRE as of 10 Jul 2024


\bibitem {RunTheta}J.~R.~Ellis and M.~K.~Gaillard,
%``Strong and Weak CP Violation,''
Nucl.\ Phys.\ B \textbf{150} (1979) 141;
%%CITATION = doi:10.1016/0550-3213(79)90297-9;%%
I.~B.~Khriplovich and A.~I.~Vainshtein, Nucl.\ Phys.\ B \textbf{414} (1994) 27
[hep-ph/9308334].
%%CITATION = doi:10.1016/0550-3213(94)90419-7;%%
, see also C.~Smith and S.~Touati,
%``Electric dipole moments with and beyond flavor invariants,''
Nucl. Phys. B \textbf{924} (2017) 417 [1707.06805],
%21 citations counted in INSPIRE as of 10 Jul 2024
and references there.

\bibitem {Shifman:2017lkj}M.~Shifman and A.~Vainshtein,
%``(In)dependence of $\Theta$ in the Higgs regime without axions,''
Mod. Phys. Lett. A \textbf{32} (2017) 14, 1750084 [1701.00467].
%8 citations counted in INSPIRE as of 05 Jul 2024


\bibitem {tHooft:1981bkw}G.~'t Hooft,
%``Topology of the Gauge Condition and New Confinement Phases in Nonabelian Gauge Theories,''
Nucl. Phys. B \textbf{190} (1981) 455.
%1683 citations counted in INSPIRE as of 10 Jul 2024


\bibitem {Lattice}C.~Alexandrou \textit{et al.},
%``Neutron electric dipole moment using lattice QCD simulations at the physical point,''
Phys. Rev. D \textbf{103} (2021) 5, 054501 [2011.01084];
%31 citations counted in INSPIRE as of 05 Jul 2024
J.~Liang \textit{et al.} [$\chi$QCD],
%``Nucleon electric dipole moment from the \ensuremath{\theta} term with lattice chiral fermions,''
Phys. Rev. D \textbf{108} (2023) 9, 094512 [2301.04331].
%14 citations counted in INSPIRE as of 05 Jul 2024


\bibitem {Ai:2020ptm}W.~Y.~Ai \textit{et al.},
%``Consequences of the order of the limit of infinite spacetime volume and the sum over topological sectors for CP violation in the strong interactions,''
Phys. Lett. B \textbf{822} (2021) 136616 [2001.07152].
%30 citations counted in INSPIRE as of 10 Jul 2024


\bibitem {KSVZ}J.~E.~Kim,
%``Weak Interaction Singlet and Strong CP Invariance,''
Phys. Rev. Lett. \textbf{43} (1979) 103;
%2738 citations counted in INSPIRE as of 31 Jan 2023
M.~A.~Shifman, A.~I.~Vainshtein and V.~I.~Zakharov,
%``Can Confinement Ensure Natural CP Invariance of Strong Interactions?,''
Nucl. Phys. B \textbf{166} (1980) 493.
%2431 citations counted in INSPIRE as of 31 Jan 2023


\bibitem {PQ}R.~D.~Peccei and H.~R.~Quinn,
%``CP Conservation in the Presence of Instantons,''
Phys. Rev. Lett. \textbf{38} (1977) 1440;
%6961 citations counted in INSPIRE as of 31 Jan 2023
%``Constraints Imposed by CP Conservation in the Presence of Instantons,''
Phys. Rev. D \textbf{16} (1977) 1791.
%3609 citations counted in INSPIRE as of 31 Jan 2023


\bibitem {Axion}S.~Weinberg,
%``A New Light Boson?,''
Phys. Rev. Lett. \textbf{40} (1978) 223;
%4896 citations counted in INSPIRE as of 31 Jan 2023
F.~Wilczek,
%``Problem of Strong  $P$  and  $T$  Invariance in the Presence of Instantons,''
Phys. Rev. Lett. \textbf{40} (1978) 279.
%4699 citations counted in INSPIRE as of 31 Jan 2023


\bibitem {DFSZ}M.~Dine, W.~Fischler and M.~Srednicki,
%``A Simple Solution to the Strong CP Problem with a Harmless Axion,''
Phys. Lett. B \textbf{104} (1981), 199-202;
%3177 citations counted in INSPIRE as of 31 Jan 2023
A.~R.~Zhitnitsky,
%``On Possible Suppression of the Axion Hadron Interactions. (In Russian),''
Sov. J. Nucl. Phys. \textbf{31} (1980) 260.
%1960 citations counted in INSPIRE as of 31 Jan 2023

\bibitem {Kamionkowski:1992mf}M.~Kamionkowski and J.~March-Russell,
%``Planck scale physics and the Peccei-Quinn mechanism,''
Phys. Lett. B \textbf{282} (1992) 137 [hep-th/9202003].
%583 citations counted in INSPIRE as of 10 Jul 2024


\bibitem {DiLuzio:2020wdo}L.~Di Luzio \textit{et al.},
%``The landscape of QCD axion models,''
Phys. Rept. \textbf{870} (2020) 1 [2003.01100].
%643 citations counted in INSPIRE as of 10 Jul 2024

\bibitem {Grenet:2021vbb}T.~Grenet \textit{et al.}
%``The Grenoble Axion Haloscope platform (GrAHal): development plan and first results,''
[2110.14406].
%38 citations counted in INSPIRE as of 10 Jul 2024

\bibitem {AxionExp}https://cajohare.github.io/AxionLimits/

\bibitem {Outside}L.~Di Luzio, F.~Mescia and E.~Nardi,
%``Redefining the Axion Window,''
Phys. Rev. Lett. \textbf{118} (2017) 3, 031801 [1610.07593].
%207 citations counted in INSPIRE as of 10 Jul 2024
T.~Higaki \textit{et al.},
%``The QCD Axion from Aligned Axions and Diphoton Excess,''
Phys. Lett. B \textbf{755} (2016) 13 [1512.05295].
%192 citations counted in INSPIRE as of 10 Jul 2024
L.~Di Luzio \textit{et al.},
%``An even lighter QCD axion,''
JHEP \textbf{05} (2021) 184 [2102.00012].
%86 citations counted in INSPIRE as of 10 Jul 2024
M.~K.~Gaillard \textit{et al.}
%``Color unified dynamical axion,''
Eur. Phys. J. C \textbf{78} (2018) 11, 972 [1805.06465].
%94 citations counted in INSPIRE as of 10 Jul 2024


\bibitem {Co:2019jts}R.~T.~Co, L.~J.~Hall and K.~Harigaya,
%``Axion Kinetic Misalignment Mechanism,''
Phys. Rev. Lett. \textbf{124} (2020) 25, 251802 [1910.14152].
%171 citations counted in INSPIRE as of 10 Jul 2024


\bibitem {Marsh:2015xka}D.~J.~E.~Marsh,
%``Axion Cosmology,''
Phys. Rept. \textbf{643} (2016) 1 [1510.07633].
%1752 citations counted in INSPIRE as of 10 Jul 2024


\bibitem {Smith:2023htu}C.~Smith,
%``On the fermionic couplings of axionic dark matter,''
Eur. Phys. J. C \textbf{84} (2024) 12 [2302.01142].
%6 citations counted in INSPIRE as of 10 Jul 2024


\bibitem {Blind}J.~Quevillon and C.~Smith,
%``Axions are blind to anomalies,''
Eur. Phys. J. C \textbf{79} (2019) 10, 822 [1903.12559];
%36 citations counted in INSPIRE as of 10 Jul 2024
J.~Quevillon, C.~Smith and P.~N.~H.~Vuong,
%``Axion effective action,''
JHEP \textbf{08} (2022) 137 [2112.00553];
%16 citations counted in INSPIRE as of 10 Jul 2024
and for an application to axion-like particles, F.~Arias-Arag\'{o}n,
J.~Quevillon and C.~Smith,
%``Axion-like ALPs,''
JHEP \textbf{03} (2023) 134 [2211.04489].
%11 citations counted in INSPIRE as of 10 Jul 2024


\bibitem {BandL}J.~Quevillon and C.~Smith,
%``Baryon and lepton number intricacies in axion models,''
Phys. Rev. D \textbf{102} (2020) 7, 075031 [2006.06778];
%10 citations counted in INSPIRE as of 10 Jul 2024
%``Variations on the SU(5) axion,''
Eur. Phys. J. Plus \textbf{137} (2022) 1, 141 [2010.13683].
%5 citations counted in INSPIRE as of 10 Jul 2024


\bibitem {Arias-Aragon:2022byr}F.~Arias-Arag\'{o}n and C.~Smith,
%``Leptoquarks, axions and the unification of B, L, and Peccei-Quinn symmetries,''
Phys. Rev. D \textbf{106} (2022) 5, 055034 [2206.09810].
%3 citations counted in INSPIRE as of 10 Jul 2024

\end{thebibliography}
\end{document}